\documentclass[mathleft
]{an}
\usepackage{graphicx}
\usepackage{times}
\begin{document}

\Pagespan{789}{}
\Yearpublication{2006}%
\Yearsubmission{2005}%
\Month{11}%
\Volume{999}%
\Issue{88}%

\title{Dynamical mixing of two stellar populations in globular clusters}

\author{T. Decressin\inst{1}\fnmsep\thanks{Corresponding author:
  \email{decressin@astro.uni-bonn.de}\newline}
\and  H. Baumgardt\inst{1}
\and P. Kroupa\inst{1}
}
\titlerunning{Dynamical evolution of two stellar populations in globular clusters}
\authorrunning{T. Decressin, H. Baumgardt \& P. Kroupa}
\institute{Argelander Institute for Astronomy (AIfA), Auf dem
                H\"ugel 71, D-53121 Bonn, Germany}

\received{~}
\accepted{~}
\publonline{later}

\keywords{globular clusters: general -- stellar dynamics}

\abstract{Stars in globular clusters (GCs) exhibit a peculiar chemical pattern
  with strong abundance variations in light elements along with a constant
  abundance in heavy elements. These abundance anomalies can be explained
  by a primordial pollution due to a first generation of fast rotating
  massive stars which released slow winds into the ISM from which a second
  generation of chemically anomalous stars can be formed. 
  In particular the observed ratio of anomalous and
  standard stars in clusters can be used to constrain the dynamical
  evolution of GCs as around 95\% of the standard stars need to be lost by
  the clusters. We show that both residual gas expulsion during the cluster
  formation and long term evolution are
  needed to achieve this ratio.}
\maketitle

\section{Abundance variations in globular clusters}

\begin{figure*}[ht]
  \centering
  \includegraphics[angle=-90,width=\textwidth]{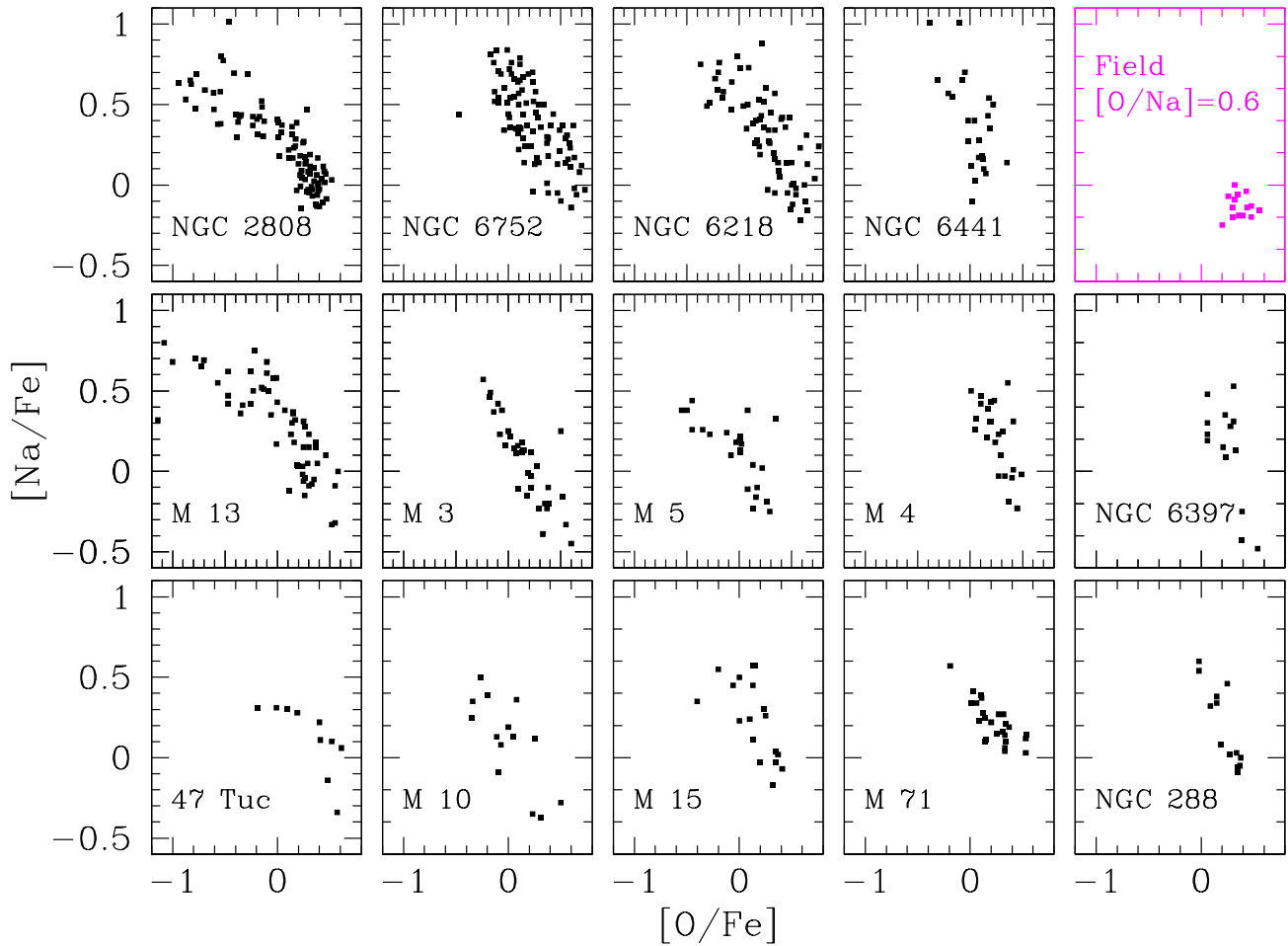}
  \caption{ONa anticorrelation
    observed in
    several clusters and in field stars. Data 
    for NGC~2808 are from Carretta et al. (2006); 
    for NGC~6752 from Carretta et al. (2007b); 
    for NGC~6218 from Carretta et al. (2007a); 
    for NGC~6441 from Gratton et al. (2007); 
    for M~3 and M~13 from Cohen \ Mel\'endez (2005) and Sneden et al. (2004); 
    for M~5 from Ivans et al. (2001); 
    for M~4 from Ivans et al. (1999); 
    for NGC~6397 from Carretta et al. (2005); 
    for 47~Tuc from Carretta et al. (2004); 
    for M~10 from Kraft et al. (1995); 
    for M~15 from Sneden et al. (1997); 
    for M~71 from Ram\'irez \& Cohen (2002); 
    for NGC~288 and NGC~362 from Shetrone \& Keane (2000); 
    and field from Gratton et al. (2000).}
  \label{fig:ONa}
\end{figure*}

It has long been known that globular cluster stars present some striking
anomalies in their content in light elements whereas their content in heavy
elements (i.e., Fe-group, $\alpha$-elements) is fairly constant from star
to star (with the notable exception of $\omega$~Cen).  While in all the
Galactic globular clusters studied so far one finds ``normal'' stars with
detailed chemical composition similar to those of field stars of same
metallicity (i.e., same [Fe/H]), one also observes numerous ``anomalous''
main sequence and red giant stars that are simultaneously deficient (to
various degrees) in C, O, and Mg, and enriched in N, Na, and Al (for recent
reviews see Gratton, Sneden \& Carretta 2004 and Charbonnel 2005). In
Fig.~\ref{fig:ONa} we present a compilation of data about the observed O-Na
anticorrelation in a sample of clusters along with metal-poor halo field
stars where such abundance patterns are not detected. Additionally, the abundance of the fragile Li was found to be
anticorrelated with that of Na in turnoff stars in a couple of globular
clusters (Pasquini et al. 2005, Bonifacio et al. 2007).

These abundance variations are expected to result from H-burning
nucleosynthesis at high temperatures up to $75\times 10^6$~K
(Denisenkov \& Denisenkova 1989, 1990, Langer \& Hoffman 1995, Prantzos et
al. 2007). Since
the low-mass stars still on the main sequence or on the RGB display such
abundance variations and since these low-mass stars do not produce the
required high temperatures to create abundance anomalies, it is expected
that they have inherited this chemical pattern at birth.

Here we follow the work of Decressin et al. (2007b) who explore the role
of a high initial rotation in such stars, finding that abundance anomalies
in globular clusters can lead to an early pollution by fast rotating
massive stars. Indeed during main sequence evolution, angular momentum is
transported from the centre to the stellar surface, and for stars heavier
than 20~$M_\odot$  with high initial rotation, their surface reaches break-up
rotational velocity at the equator (i.e., the centrifugal equatorial force
balances gravity). In such a situation, a slow mechanical wind develops at
the equator and forms a disc around the stars similar to what happens to Be
stars (Townsend et al. 2004, Ekstrom et al. 2008). The second effect of
rotation is to transport elements from the convective H-burning core to the
stellar surface which enriches the disc with H-burning material. While
matter in discs has a slow escape velocity and hence will stay in the
potential well of the cluster, matter released during the main part of the
He-burning phase and during SN explosions has a very high radial velocity
and will be lost by the cluster. Therefore, new stars can only form from
the matter available in discs and can become the stars with abundance
anomalies we observe today. Thus globular clusters can contain two 
populations of low-mass stars: a first generation which has the chemical
composition of the material out of which the cluster formed (similar to
field stars with similar metallicity); and a second generation of stars
harbouring the abundance anomalies born from the ejecta of fast rotating
massive stars.

\section{Number ratio between two populations in globular clusters}

Based on the determination of the composition of giant stars in NGC~6752
by Carretta et al. (2007b), Decressin, Charbnnel \& Meynet (2007a) determined
that around 85\% of stars (of the sample of 120 stars) display abundance
anomalies. Prantzos \& Charbonnel (2006) find similar results for NGC 2808
with their analysis of the data of Carretta et al. (2006): 70\% of
the cluster stars present abundance anomalies.

To produce such a high fraction of chemically peculiar stars, the main
problem is that assuming a Salpeter (1955) IMF for the polluters, the
accumulated mass of the slow winds ejected by the fast rotating massive
stars would only form 
10\% of the total number of the low-mass stars. To match the observations, one thus requires either (a) a flat
IMF with a slope of 0.55 instead of 1.35 (Salpeter value), or (b) that 95\%
of the first generation unpolluted stars have escaped the cluster
Decressin et al. (2007a). In
this paper we want to quantify if such high loss of stars is possible and
which are the main processes which drive it.

\section{Dynamical evolution of globular clusters}

\begin{figure}[ht]
  \centering
  \includegraphics[width=0.48\textwidth]{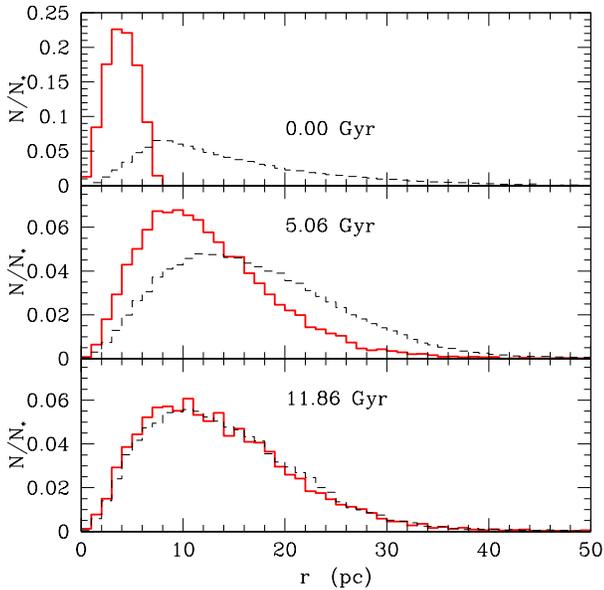}
  \caption{Radial distribution of the first (dashed lines) and second
    (full lines) generation of low-mass
    stars at three different times. Each histogram is
    normalised to the total number of stars in each population.}
  \label{fig:distr}
\end{figure}

\begin{figure}[ht]
  \centering
  \includegraphics[width=0.48\textwidth]{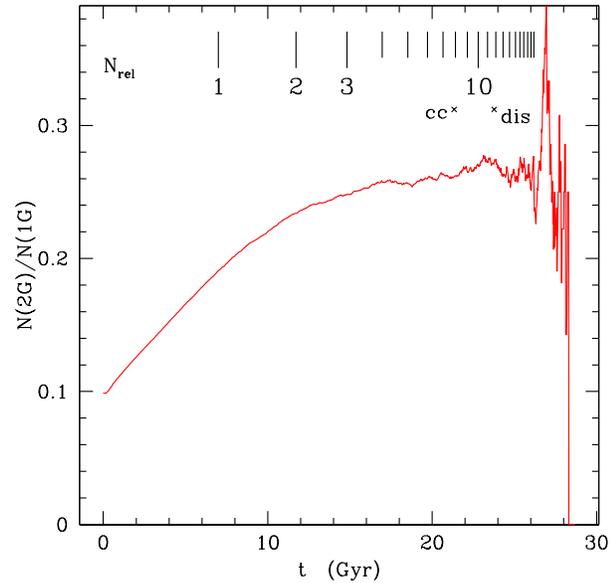}
  \caption{Number ratio between the second (with low initial specific
    energy) and first (with high initial specific energy) population of
    low-mass stars in a cluster with initially 128k stars as a function of
    time. At the top of each panel the number of passed relaxation times is
    shown, crosses indicate the time of core
      collapse and of cluster dissolution.}
  \label{fig:ratio}
\end{figure}

First we assume that the globular clusters display primordial mass
segregation. Thus the massive stars are located in the center of the
clusters. As we expect that the formation of the second generation of
low-mass stars happens locally around each massive star (see
Decressin et al. 2007a for more details), the second generation
of stars will also be more centrally concentrated than the first
generation. 
In such a situation, two competitive processes act in
the clusters: the loss of stars in the outer cluster parts will first
reduce the number of bound first generation stars; and the dynamical spread
of the initially more concentrated second generation stars will stop this
differential loss when the two populations are dynamically mixed. 

Our analysis 
based on the N-body models computed by Baumgardt \& Makino (2003) with the
collisional Aarseth N-body code \textsc{nbody4} (Aarseth 1999) is presented
in details in Decressin, Baumgardt \& Kroupa (2008).

As these models have been computed for a
single stellar population, we apply the following process to mimic the
formation of a cluster with two dynamically distinct
populations: we sort all the low-mass
stars ($M\le 0.9$~$M_\odot$) according to their specific energy (i.e., their
energy per unit mass). We define the
second stellar generation as the stars with lowest specific
energies,
(i.e., those which are most tightly bound to the cluster due to their small
central distance and low velocity). The number of second generation stars
is given by having their total number represent 10\% of the total number of
low-mass stars.

In Fig.~\ref{fig:distr} one
can see the radial distribution of the two populations at the same
epochs. Initially, the second generation stars with low specific energy are
concentrated within 6~pc around the centre of the cluster while stars of
the first generation show a more extended distribution up to
40~pc.

Progressively the second generation stars spread out due to dynamical
encounters so their radial distribution extends. The middle panel in
Fig.~\ref{fig:distr} shows that even after 5~Gyr of
evolution the two populations have still different distributions. 
The bottom panel of \ref{fig:distr} shows that after
nearly 12~Gyr of evolution (slightly more than 2 initial relaxation times) the
two populations have similar radial distributions and can no longer be
distinguished owing to their dynamical properties.

As previously seen, the effect of the external potential of the Galaxy on
the cluster is to strip away stars lying in the outer part of the
cluster. 
Initially, only stars of the first generation populate the outer
part of the cluster owing to their high specific energy. Therefore only
first generation stars are lost in the beginning. This lasts until the
second generation stars migrate into the outer part of the
cluster. Depending on the cluster mass, it takes between 1 to 4~Gyr to
start a loss of second generation stars. 
Due to the time-delay to lose second generation stars, their remaining
fraction in the cluster is always higher than that of the first generation
stars except during the final stage of cluster dissolution.
Fig.~\ref{fig:ratio} quantifies this point by showing the time evolution
of the number ratio of second to first generation stars. As a direct
consequence of our selection procedure, the initial ratio is 0.1; and it then
increases gradually with time. It tends to stay nearly constant as
soon as the two distributions are similar. Finally, at the time of 
cluster dissolution (i.e., when the cluster has lost 95\% of its initial
mass, indicated by the label ``dis'' in Fig.~\ref{fig:distr}), large variations occur due to the low number of
low-mass stars present in the cluster. In Fig.\ref{fig:distr} we have also
indicated the
number of passed relaxation times, showing that the increase of the number
ratio last only 3 relaxation time.

Over the cluster history, the fraction of second generation stars relative
to first generation ones increases by a factor of 2.5. Therefore, when the
two populations have the same radial distribution, these stars represent
25\% of the low-mass stars present in the clusters. Compared to the observed
ratios (70\% and 85\% in NGC~2808 and NGC~6752 respectively)
the internal
dynamical evolution and the dissolution due to the tidal forces of the host
Galaxy are not efficient enough. An
additional mechanism is thus needed to expel the first generation stars
more effectively.

\section{Gas expulsion}

\begin{figure}
  \includegraphics[width=0.48\textwidth]{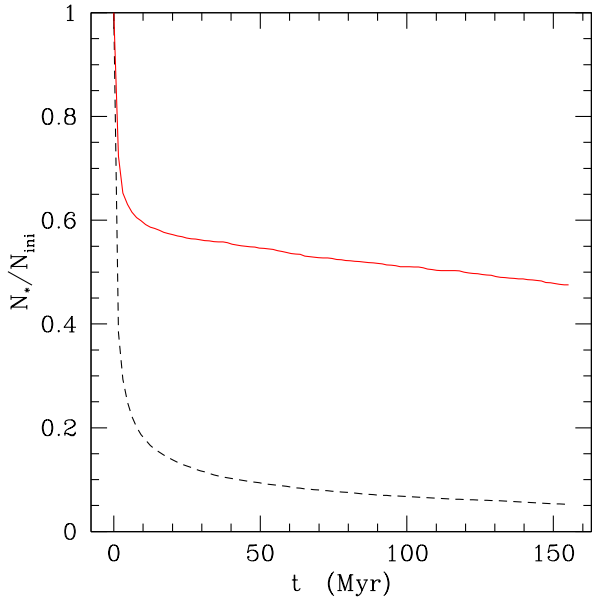}
  \caption{Evolution of the number of first (dashed line) and second (full
    line) generation stars still bound to the cluster with
  initial parameters: SFE of 0.33, $r_h/r_t = 0.06$ and
  $\tau_M/t_{\rm{Cross}} = 0.33$.} 
  \label{fig:GE}
\end{figure}

Initial gas expulsion by supernovae is an ideal candidate 
as it operates early in the cluster history (a few million
years after cluster formation). As the gas still present after the star
formation is removed,
the potential well of the cluster can be strongly reduced and the outer
parts of the cluster can become unbound. 

This process has been recently investigated by Baumgardt \& Kroupa (2007)
who have performed a complete grid of N-body models to study its influences
on the cluster evolution by varying the free parameters: star formation
efficiency, SFE, ratio between the half-mass and tidal radius, $r_h/r_t$,
and the ratio between the timescale for gas expulsion to the crossing time,
$\tau_M/t_{\rm{Cross}}$.  In particular they show that gas expulsion can
lead to the complete disruption of the cluster in some extreme cases.  It
should be noted that this process has already been used successfully by
Marks, Kroupa, Baumgardt (2008) to explain the challenging correlation
between the central concentration and the mass function of globular
clusters as found by DeMarchi, Paresce \&  Pulone (2007).
 
We have repeated the
same method to obtain two populations in globular clusters as we show in
\S~3 to the models of Baumgardt \& Kroupa (2007). Fig.~\ref{fig:GE} shows
that 
in the case of clusters which lose around 90\% of
their stars, most first generation stars are lost. At the end of the
computation only 5\% of first generation stars remain bound to the cluster
and around half of second
generation stars also escape the cluster. Therefore the number ratio between
the second to first generation stars can only increase from 0.1 to 1:
half of the low-mass stars are second generation stars. It should be
noted that the radial distribution of both populations differs, the second
generation stars being more centrally distributed, so that we
can expect that this ratio will increase in the long-term evolution of the
cluster (see Decressin et al., in preparation).

\acknowledgements

T. D. acknowledges financial support from swiss FNS.




\begin{thebibliography}{35}
\expandafter\ifx\csname natexlab\endcsname\relax\def\natexlab#1{#1}\fi

\bibitem[{{Aarseth}(1999)}]{Aarseth1999}
{Aarseth}, S.~J.: 1999, \pasp 111, 1333

\bibitem[{{Baumgardt} \& {Kroupa}(2007)}]{BaumgardtKroupa2007}
{Baumgardt}, H. \& {Kroupa}, P.: 2007, \mnras 380, 1589

\bibitem[{{Baumgardt} \& {Makino}(2003)}]{BaumgardtMakino2003}
{Baumgardt}, H. \& {Makino}, J.: 2003, \mnras 340, 227

\bibitem[{{Bonifacio} {et~al.}(2007){Bonifacio}, {Pasquini}, {Molaro},
  {Carretta}, {Fran{\c c}ois}, {Gratton}, {James}, {Sbordone}, {Spite}, \&
  {Zoccali}}]{BonifacioPasquini2007}
{Bonifacio}, P., {Pasquini}, L., {Molaro}, P., {et~al.}: 2007, A\&A 470, 153

\bibitem[{{Carretta} {et~al.}(2007{\natexlab{a}}){Carretta}, {Bragaglia},
  {Gratton}, {Catanzaro}, {Leone}, {Sabbi}, {Cassisi}, {Claudi}, {D'Antona},
  {Fran{\c c}ois}, {James}, \& {Piotto}}]{CarrettaBragaglia2007b}
{Carretta}, E., {Bragaglia}, A., {Gratton}, R.~G., {et~al.}: 2007{\natexlab{a}},
  A\&A 464, 939

\bibitem[{{Carretta} {et~al.}(2006){Carretta}, {Bragaglia}, {Gratton}, {Leone},
  {Recio-Blanco}, \& {Lucatello}}]{CarrettaBragaglia2006}
{Carretta}, E., {Bragaglia}, A., {Gratton}, R.~G., {et~al.}: 2006, A\&A 450,
  523

\bibitem[{{Carretta} {et~al.}(2007{\natexlab{b}}){Carretta}, {Bragaglia},
  {Gratton}, {Lucatello}, \& {Momany}}]{CarrettaBragaglia2007}
{Carretta}, E., {Bragaglia}, A., {Gratton}, R.~G., {Lucatello}, S., \&
  {Momany}, Y.: 2007{\natexlab{b}}, A\&A 464, 927

\bibitem[{{Carretta} {et~al.}(2004){Carretta}, {Gratton}, {Bragaglia},
  {Bonifacio}, \& {Pasquini}}]{CarrettaGratton2004}
{Carretta}, E., {Gratton}, R.~G., {Bragaglia}, A., {Bonifacio}, P., \&
  {Pasquini}, L.: 2004, A\&A, 416 925

\bibitem[{{Carretta} {et~al.}(2005){Carretta}, {Gratton}, {Lucatello},
  {Bragaglia}, \& {Bonifacio}}]{CarrettaGratton2005}
{Carretta}, E., {Gratton}, R.~G., {Lucatello}, S., {Bragaglia}, A., \&
  {Bonifacio}, P.: 2005, A\&A 433, 597

\bibitem[{{Charbonnel}(2005)}]{Charbonnel2005}
{Charbonnel}, C.: 2005, in IAU Symposium, ed. V.~{Hill}, P.~{Fran{\c c}ois}, \&
  F.~{Primas} 347--356

\bibitem[{{Cohen} \& {Mel{\'e}ndez}(2005)}]{Cohen2005}
{Cohen}, J.~G. \& {Mel{\'e}ndez}, J.: 2005, \aj 129, 303

\bibitem[{{De Marchi} {et~al.}(2007){De Marchi}, {Paresce}, \&
  {Pulone}}]{DeMarchiParesce2007}
{De Marchi}, G., {Paresce}, F., \& {Pulone}, L.: 2007, \apjl 656, L65

\bibitem[{{Decressin} {et~al.}(2008){Decressin}, {Baumgardt}, \&
  {Kroupa}}]{DecressinBaumgardt2008}
{Decressin}, T., {Baumgardt}, H., \& {Kroupa}, P.: 2008, A\&A submitted

\bibitem[{{Decressin} {et~al.}(2007{\natexlab{a}}){Decressin}, {Charbonnel}, \&
  {Meynet}}]{DecressinCharbnnel2007}
{Decressin}, T., {Charbonnel}, C., \& {Meynet}, G.: 2007{\natexlab{a}}, A\&A
  475, 859

\bibitem[{{Decressin} {et~al.}(2007{\natexlab{b}}){Decressin}, {Meynet},
  {Charbonnel}, {Prantzos}, \& {Ekstr{\"o}m}}]{DecressinMeynet2007}
{Decressin}, T., {Meynet}, G., {Charbonnel}, C., {Prantzos}, N., \&
  {Ekstr{\"o}m}, S.: 2007{\natexlab{b}}, A\&A 464, 1029

\bibitem[{{Denisenkov} \& {Denisenkova}(1989)}]{DenisenkovDenisenkova1989}
{Denisenkov}, P.~A. \& {Denisenkova}, S.~N.: 1989, Astronomicheskij Tsirkulyar
  1538, 11

\bibitem[{{Denisenkov} \& {Denisenkova}(1990)}]{DenisenkovDenisenkova1990}
{Denisenkov}, P.~A. \& {Denisenkova}, S.~N.: 1990, Soviet Astronomy Letters 16,
  275

\bibitem[{{Ekstr{\"o}m} {et~al.}(2008){Ekstr{\"o}m}, {Meynet}, {Maeder}, \&
  {Barblan}}]{EkstromMeynet2008}
{Ekstr{\"o}m}, S., {Meynet}, G., {Maeder}, A., \& {Barblan}, F.: 2008, A\&A
  478, 467

\bibitem[{{Gratton} {et~al.}(2004){Gratton}, {Sneden}, \&
  {Carretta}}]{GrattonSneden2004}
{Gratton}, R., {Sneden}, C., \& {Carretta}, E.: 2004, \araa 42, 385

\bibitem[{{Gratton} {et~al.}(2007){Gratton}, {Lucatello}, {Bragaglia},
  {Carretta}, {Cassisi}, {Momany}, {Pancino}, {Valenti}, {Caloi}, {Claudi},
  {D'Antona}, {Desidera}, {Fran{\c c}ois}, {James}, {Moehler}, {Ortolani},
  {Pasquini}, {Piotto}, \& {Recio-Blanco}}]{GrattonLucatello2007}
{Gratton}, R.~G., {Lucatello}, S., {Bragaglia}, A., {et~al.}: 2007, A\&A 464,
  953

\bibitem[{{Gratton} {et~al.}(2000){Gratton}, {Sneden}, {Carretta}, \&
  {Bragaglia}}]{GrattonSneden2000}
{Gratton}, R.~G., {Sneden}, C., {Carretta}, E., \& {Bragaglia}, A.: 2000, A\&A
  354, 169

\bibitem[{{Ivans} {et~al.}(2001){Ivans}, {Kraft}, {Sneden}, {Smith}, {Rich}, \&
  {Shetrone}}]{IvansKraft2001}
{Ivans}, I.~I., {Kraft}, R.~P., {Sneden}, C., {et~al.}: 2001, \aj 122, 1438

\bibitem[{{Ivans} {et~al.}(1999){Ivans}, {Sneden}, {Kraft}, {Suntzeff},
  {Smith}, {Langer}, \& {Fulbright}}]{IvansSneden1999}
{Ivans}, I.~I., {Sneden}, C., {Kraft}, R.~P., {et~al.}: 1999, \aj 118, 1273

\bibitem[{{Kraft} {et~al.}(1995){Kraft}, {Sneden}, {Langer}, {Shetrone}, \&
  {Bolte}}]{KraftSneden1995}
{Kraft}, R.~P., {Sneden}, C., {Langer}, G.~E., {Shetrone}, M.~D., \& {Bolte},
  M.: 1995, \aj 109, 2586

\bibitem[{{Langer} \& {Hoffman}(1995)}]{LangerHoffman1995}
{Langer}, G.~E. \& {Hoffman}, R.~D.: 1995, \pasp 107, 1177

\bibitem[{{Marks} {et~al.}(2008){Marks}, {Kroupa}, \&
  {Baumgardt}}]{MarksKroupa2008}
{Marks}, M., {Kroupa}, P., \& {Baumgardt}, H.: 2008, MNRAS 386, 2047

\bibitem[{{Pasquini} {et~al.}(2005){Pasquini}, {Bonifacio}, {Molaro},
  {Francois}, {Spite}, {Gratton}, {Carretta}, \&
  {Wolff}}]{PasquiniBonifacio2005}
{Pasquini}, L., {Bonifacio}, P., {Molaro}, P., {et~al.}: 2005, A\&A 441, 549

\bibitem[{{Prantzos} \& {Charbonnel}(2006)}]{PrantzosCharbonnel2006}
{Prantzos}, N. \& {Charbonnel}, C.: 2006, A\&A 458, 135

\bibitem[{{Prantzos} {et~al.}(2007){Prantzos}, {Charbonnel}, \&
  {Iliadis}}]{PrantzosCharbonnel2007}
{Prantzos}, N., {Charbonnel}, C., \& {Iliadis}, C.: 2007, A\&A 470, 179

\bibitem[{{Ram{\'{\i}}rez} \& {Cohen}(2002)}]{RamirezCohen2002}
{Ram{\'{\i}}rez}, S.~V. \& {Cohen}, J.~G.: 2002, \aj 123, 3277

\bibitem[{{Salpeter}(1955)}]{Salpeter1955}
{Salpeter}, E.~E.: 1955, \apj 121, 161

\bibitem[{{Shetrone} \& {Keane}(2000)}]{ShetroneKeane2000}
{Shetrone}, M.~D. \& {Keane}, M.~J.: 2000, \aj 119, 840

\bibitem[{{Sneden} {et~al.}(2004){Sneden}, {Kraft}, {Guhathakurta}, {Peterson},
  \& {Fulbright}}]{SnedenKraft2004}
{Sneden}, C., {Kraft}, R.~P., {Guhathakurta}, P., {Peterson}, R.~C., \&
  {Fulbright}, J.~P.: 2004, \aj 127, 2162

\bibitem[{{Sneden} {et~al.}(1997){Sneden}, {Kraft}, {Shetrone}, {Smith},
  {Langer}, \& {Prosser}}]{SnedenKraft1997}
{Sneden}, C., {Kraft}, R.~P., {Shetrone}, M.~D., {et~al.}: 1997, \aj 114, 1964

\bibitem[{{Townsend} {et~al.}(2004){Townsend}, {Owocki}, \&
  {Howarth}}]{TownsendOwocki2004}
{Townsend}, R.~H.~D., {Owocki}, S.~P., \& {Howarth}, I.~D.: 2004, \mnras 350,
  189

\end{thebibliography}

\end{document}